\documentclass[aps,pra,reprint,groupedaddress]{revtex4-2}

\usepackage{xcolor}
\usepackage{hyperref}
\usepackage{amsmath}
\usepackage{amsfonts}
\usepackage{amssymb}
\usepackage{graphicx}
\usepackage{empheq}
\usepackage{orcidlink}
\usepackage{enumitem}
\usepackage{lipsum}

\definecolor{linkcolour}{HTML}{000066}	%light purple link for the email
\hypersetup{colorlinks,breaklinks,
	urlcolor=linkcolour, 
	linkcolor=linkcolour,
	citecolor=linkcolour}

\begin{document}

\title{Role of Spatial Coherence in Single-Shot Lensless Image Reconstruction}

\author{Ganesh M. Balasubramaniam$^1$\orcidlink{0000-0001-7824-5831}, Xiao-Liu Chu$^{2}$\orcidlink{0000-0002-7476-2097} and Matthew R. Foreman$^{1,2}$\orcidlink{0000-0001-5864-9636}}
\email[]{matthew.foreman@ntu.edu.sg}
\affiliation{
$^1$School of Electrical and Electronic Engineering, Nanyang Technological University, 50 Nanyang Avenue, Singapore 639798 \\
$^2$Institute for Digital Molecular Analytics and Science, 59 Nanyang Drive, Singapore 636921
}

\date{\today}

\begin{abstract}
Lensless imaging is a technique that recovers object information computationally from diffraction patterns recorded without imaging optics, making its performance strongly dependent on the forward model used during image reconstruction. A common source of reconstruction error is the assumption of fully coherent propagation, even when the illumination exhibits partial spatial coherence in practice. Here, the role of spatial coherence is examined for assorted object classes using a coherence-aware forward model based on generalised van Cittert-Zernike Schell propagation. Simulated measurements are generated over a controlled range of effective source spatial coherence lengths and reconstructed using either a partially coherent forward model or a conventional coherent propagation model. Our results show that decreasing spatial coherence progressively degrades reconstructions under the coherent assumption, whereas incorporating partial coherence can preserve object structure and improve image quality, particularly for dense high-spatial-frequency features. Reconstructions from experimental measurements further confirm the practical need to use coherence-aware inversion under partially spatially coherent illumination. These findings establish spatial coherence as a defining component of the inverse problem in lensless imaging and provide a route to optimization of partially coherent lensless imaging systems.
\end{abstract}

\maketitle

Lensless imaging provides an alternative to conventional microscopy by computationally recovering object information from diffraction patterns recorded directly on an image sensor. By eliminating imaging optics, lensless systems combine a compact form factor with a large space-bandwidth product, making them attractive for wide-field microscopy, high-throughput screening, and portable imaging platforms \cite{Bishara2010WideFOV,Greenbaum2012WideField,Balasubramaniam2026SingleShotGP}. In such systems, the detector does not record an image of the specimen in the conventional sense. Instead, a diffraction-encoded intensity pattern is measured, from which the object must be estimated. In model-based reconstruction algorithms, a forward model maps a candidate object to a detector intensity under the assumed illumination and propagation conditions. Consequently, reconstruction quality is determined not only by the numerical formulation of the inverse problem, but also by the physical accuracy of the forward model.

A key component of a forward model is the spatial coherence of the illumination. Coherent propagation models are widely used because they yield a simple wave-optical description of diffraction and integrate readily into inverse reconstruction algorithms. Many practical lensless imaging systems, however, operate with extended sources, such as light-emitting diodes, for which the object plane illumination is only partially spatially coherent \cite{Isikman2011Tomographic,Chen2022SingleShot}. Partial coherence is not a minor perturbation to an optical system.  It is known to influence resolution, contrast transfer, and reconstruction performance across tomographic microscopy, high-numerical-aperture focusing, and quantitative phase retrieval \cite{Isikman2011Tomographic,Chen2022SingleShot, Foreman2009FocusingPCPolarized}. Fundamentally, the transverse extent of the source determines the mutual coherence of the incident field, which in turn modifies interference visibility, diffraction contrast, and the spatial structure of the intensity at the detector, thus affecting lensless image formation \cite{Bian2019Designing,Cai2007LenslessPC}. Since single-shot reconstruction relies entirely on the assumed forward model, these coherence-dependent changes in the diffraction pattern are incorrectly attributed to the object, erasing fine features or introducing artefacts.  Whether coherent propagation suffices for a given object, or whether coherence-aware inversion is required, therefore remains an open question whose resolution necessitates isolating coherence-induced error from that associated with sampling, regularisation, and numerical optimisation.

In this work, coherence-induced model mismatch in lensless imaging is examined directly using a forward model based on the generalised van Cittert-Zernike Schell propagator, which provides an efficient description of the field diffracted by an object under partially spatially coherent illumination \cite{Montoya2025GVS}. Lensless measurements are simulated across a controlled range of source radii for both amplitude and phase objects, and reconstructions obtained with a coherence-aware inverse model are compared with those found using a conventional coherent propagator. Our analysis establishes how decreasing spatial coherence alters the sensor measurement, quantifies the resulting degradation when coherence is neglected during inversion, and demonstrates that explicitly incorporating the illumination coherence can restore reconstruction quality across distinct object classes. We further support our findings through an experimental demonstration.

\begin{figure}[t]
	\centering
	\includegraphics[width=\linewidth]{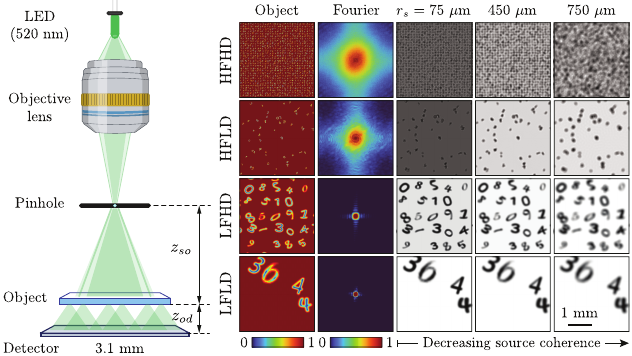}
	\caption{\textbf{Coherence dependent lensless image formation.}
		(left) Lensless imaging geometry, where source spatial coherence is controlled by the pinhole diameter. 
		(right) Columns, from left to right, show the object, corresponding normalised log Fourier spectrum \(\widetilde{S}_{O}(f_x,f_y)\), and simulated detector intensities for \(r_s=75~\mu\mathrm{m}\), \(450~\mu\mathrm{m}\), and \(750~\mu\mathrm{m}\) respectively for different object classes (top to bottom): high-frequency high-density (HFHD), high-frequency low-density (HFLD), low-frequency high-density (LFHD), and low-frequency low-density (LFLD).}
	\label{fig:coherence_forward_model}
\end{figure}

We consider a lensless forward imaging model defined for a thin transmissive object with complex transmittance
\begin{equation}
	t(\mathbf{r}_{o})
	=
	a(\mathbf{r}_{o})
	\exp\!\left[
	i\phi(\mathbf{r}_{o})
	\right],
	\
	0 \leq a(\mathbf{r}_{o}) \leq 1,
	\
	0 \leq \phi(\mathbf{r}_{o}) \leq 2\pi ,
	\label{eq:object_transmittance}
\end{equation}
where \(\mathbf{r}_{o}=(x_{o},y_{o})\) is the transverse coordinate in the object plane, \(a(\mathbf{r}_{o})\) is the amplitude transmittance,  and $\phi(\mathbf{r}_o)$ is the object-induced phase delay. The source plane, object plane, and detector plane are separated by distances \(z_{so}\) and \(z_{od}\), as shown in Fig.~\ref{fig:coherence_forward_model}. When the object is illuminated by a coherent spherical wave, the detector-plane intensity is given by
\begin{equation}
	I_{\mathrm{coh}}(\mathbf{r}_{d};z_{od})
	=
	\left|
	\mathcal{P}_{z_{od}}
	\left\{
	t(\mathbf{r}_{o})
	\exp\!\left[
	\frac{i k}{2z_{so}}
	|\mathbf{r}_{o}|^{2}
	\right]
	\right\}
	(\mathbf{r}_{d})
	\right|^{2},
	\label{eq:coherent_forward_model}
\end{equation}
where \(\mathbf{r}_{d}=(x_{d},y_{d})\) denotes the transverse detector coordinate, \(k=2\pi/\lambda\) is the wavenumber, and \(\mathcal{P}_{z_{od}}\{\cdot\}\) represents free-space propagation from the object plane to the detector plane, which we implement using the angular spectrum method.

Partial spatial coherence of the illumination source can additionally be incorporated using the generalised van Cittert-Zernike Schell formalism \cite{Montoya2025GVS}. The normalised coherence function is a function of the separation between two points in the object plane, \(\Delta \mathbf{r}_{o}=(\Delta x_{o},\Delta y_{o})\) and is given by
\begin{equation}
	\mu_{F}(\boldsymbol{\xi}_{o})
	=
	\frac{
		\mathcal{F}_{\mathbf{r}_{s}\rightarrow \boldsymbol{\xi}_{o}}
		\left\{
		S(\mathbf{r}_{s})
		\right\}
	}{
		\left|
		\mathcal{F}_{\mathbf{r}_{s}\rightarrow \boldsymbol{\xi}_{o}}
		\left\{
		S(\mathbf{r}_{s})
		\right\}
		\right|_{\max}
	},
	\qquad
	\boldsymbol{\xi}_{o}
	\triangleq
	\frac{\Delta \mathbf{r}_{o}}{\lambda z_{so}},
	\label{eq:complex_coherence_factor}
\end{equation}
where \(S(\mathbf{r}_{s})\) is the intensity distribution at the pinhole source (see Fig.~\ref{fig:coherence_forward_model}) and \(\mathcal{F}_{\mathbf{r}_{s}\rightarrow \boldsymbol{\xi}_{o}}\{\cdot\}\) denotes the two-dimensional Fourier transform from the source-plane coordinate \(\mathbf{r}_{s}=(x_{s},y_{s})\) to the spatial-frequency coordinate \(\boldsymbol{\xi}_{o}\). In this case, the detector-plane intensity  (neglecting an unimportant constant scaling factor) is
\begin{equation}
	I_{\mathrm{pc}}(\mathbf{r}_{d};z_{od}) =
	\mathcal{F}_{\Delta \mathbf{r}_{o}\rightarrow \mathbf{r}_{d}}
	\left[
	\mathcal{F}^{-1}_{\mathbf{r}_{d}\rightarrow \Delta \mathbf{r}_{o}}
	\left[
	I_{\mathrm{coh}}(\mathbf{r}_{d};z_{od})
	\right]    \mu_{F}
	\left(\boldsymbol{\xi}_o
	\right)
	\right].
	\label{eq:partially_coherent_forward_model}
\end{equation}
The source distribution is taken here to be uniform, i.e., \(S(\mathbf{r}_{s})=\mathrm{circ}(|\mathbf{r}_{s}|/r_s)\), and the pinhole radius \(r_s\) is varied to control the spatial coherence of the illumination. In the fully coherent point-source limit, \(\mu_F(\boldsymbol{\xi}_{o})=1\), such that \eqref{eq:partially_coherent_forward_model} reduces to \eqref{eq:coherent_forward_model}. 

To examine coherence-dependent image formation, we simulated ensembles of amplitude-only objects ($\phi(\mathbf{r}_o)=0$) which vary in two structural properties, namely the spatial frequency content of \(a(\mathbf{r}_{o})\) and the real-space feature density (see Supplementary Material).  Specifically, these span high-frequency high-density (HFHD), high-frequency low-density (HFLD), low-frequency high-density (LFHD), and low-frequency low-density (LFLD) object classes. 
Representative examples from each class are shown in Fig.~\ref{fig:coherence_forward_model} (first column). The second column depicts the corresponding normalised log Fourier spectrum,
\begin{equation}
	\widetilde{S}_{O}(f_x,f_y)
	=
	\mathcal{N}
	\left[
	\log
	\left\{
	1+
	\left|
	\mathcal{F}\{a(\mathbf{r}_{o})\}
	\right|
	\right\}
	\right],
	\label{eq:normalised_log_fourier_spectrum}
\end{equation}
where \(f_x\) and \(f_y\) are the horizontal and vertical spatial frequencies, and \(\mathcal{N}[\cdot]\) (or equivalently \(\widetilde{[\cdot]}\)) linearly maps the input function to \([0,1]\). For a fixed object and propagation geometry, increasing the source radius modifies the detector-plane intensity through the coherence factor in \eqref{eq:complex_coherence_factor}. The resulting recorded lensless measurement is seen to depend directly on the source's spatial coherence, as shown in the final three columns of Fig.~\ref{fig:coherence_forward_model}. Corresponding results for phase-objects (whereby $a(\mathbf{r}_o)=1$, and $a(\mathbf{r}_o) \rightarrow \phi(\mathbf{r}_o)$ in \eqref{eq:normalised_log_fourier_spectrum}) are also presented in the Supplementary Material.

The effect of coherence-induced model mismatch during inversion was subsequently assessed by reconstructing each object from the simulated detector-plane intensity assuming either a coherence-aware operator, \(\mathcal{M}_{\mathrm{pc}}\), or a fully coherent operator \(\mathcal{M}_{\mathrm{coh}}\), which evaluate \eqref{eq:partially_coherent_forward_model} and \eqref{eq:coherent_forward_model} respectively. For a measured detector intensity \(I_{\mathrm{meas}}\), the reconstructed object associated with either operator is obtained from
\begin{equation}
	\widehat{t}_{\nu}
	=
	\underset{t\in\mathcal{C}}{\operatorname*{arg\,min}}
	\;
	\mathcal{J}_{\nu}(t),
	\qquad
	\nu\in
	\left\{
	\mathrm{pc},
	\mathrm{coh}
	\right\},
	\label{eq:inverse_problem}
\end{equation}
where \(\mathcal{C}\) denotes the object constraint set. The reconstruction objective is defined as
\begin{equation}
	\begin{aligned}
		&\mathcal{J}_{\nu}(t)
		=
		w_{\mathrm{SSIM}}
		\left[
		1-
		\operatorname{SSIM}
		\left(
		\widetilde{\mathcal{M}_{\nu}\{t\}},
		\widetilde{I}_{\mathrm{meas}}
		\right)
		\right]
		\\
		&\quad+
		w_{\mathrm{MSE}}
		\left\|
		\widetilde{\mathcal{M}_{\nu}\{t\}}
		-
		\widetilde{I}_{\mathrm{meas}}
		\right\|_{2}^{2}
		+
		\lambda_{\mathrm{TV}}
		\mathcal{R}_{\mathrm{TV}}(t)
		+
		\lambda_{\mathrm{bg}}
		\mathcal{R}_{\mathrm{bg}}(t),
	\end{aligned}
	\label{eq:reconstruction_objective}
\end{equation}
where $\mathcal{R}_{\mathrm{TV}}(t)
=
\sum_{\mathbf{r}}
[
\left|\partial_x t(\mathbf{r})\right|^2
+
\left|\partial_y t(\mathbf{r})\right|^2
+
\epsilon_{\mathrm{TV}}^2
]^{1/2}$ is an isotropic total variation penalty that suppresses spurious local oscillations while preserving sharp object transitions, and $\mathcal{R}_{\mathrm{bg}}(t)
=
\left\|
B_t(\mathbf{r})
\odot
\left[
a(\mathbf{r})
-
\left(\mathcal{G}_{\sigma} * a\right)(\mathbf{r})
\right]
\right\|_2^2$ suppresses high-spatial-frequency background fluctuations without strongly penalising object boundaries. 
The map \(B_t(\mathbf{r})\) is a self-estimated background weight computed from the current object estimate. Pixels with weak object contrast and low local gradient are assigned high weights, whereas pixels containing object structure or sharp boundaries are assigned low weights. The Gaussian kernel \(\mathcal{G}_{\sigma}\) defines the local smooth background estimate, SSIM is the structural similarity, \(*\) denotes convolution, and \(\odot\) denotes pointwise multiplication. The parameter \(\epsilon_{\mathrm{TV}}=10^{-8}\) ensures differentiability of the total variation term, while \(\lambda_{\mathrm{TV}}=10^{-4}\) and \(\lambda_{\mathrm{bg}}=10^{-5}\) set the strengths of the total variation and background regularisers respectively. \(w_{\mathrm{SSIM}}=0.3\) and \(w_{\mathrm{MSE}}=0.7\) dictate the relative weights assigned to the structural-similarity and mean-squared-error terms. The same objective, optimisation schedule, and initialisation are used for both \(\mathcal{M}_{\mathrm{pc}}\) and \(\mathcal{M}_{\mathrm{coh}}\), ensuring that any difference in reconstruction quality arises solely from the forward model used.

\begin{figure}[t]
	\centering
	\includegraphics[width=\linewidth]{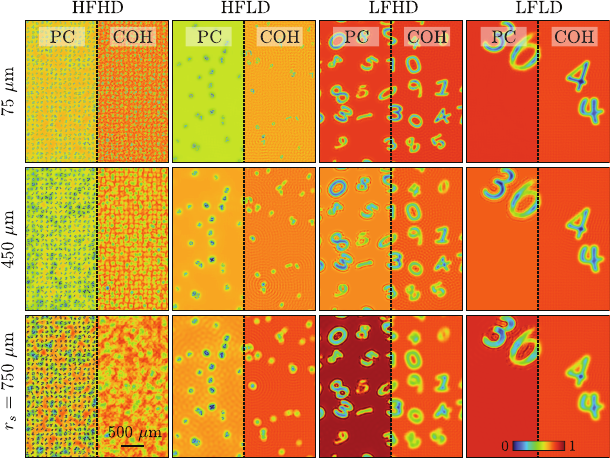}
	\caption{\textbf{Representative reconstructions across object classes and source coherence.}
		Object reconstructions for HFHD, HFLD, LFHD, and LFLD object classes (left to right) and source radii (top to bottom). In each image, the left half shows the reconstruction obtained using the partially coherent (PC) forward model, while the right half shows the reconstruction obtained using the coherent (COH) forward model. }
	\label{fig:representative_reconstructions}
\end{figure}

Reconstruction was performed from a single detector intensity recorded at \(z_{od}=2~\mathrm{mm}\), under the constraint \(0\leq a(\mathbf{r}_{o})\leq1\). For all simulated measurements, we assumed a wavelength of \(520~\mathrm{nm}\), \(z_{so}=3~\mathrm{cm}\), whilst \(r_s\) was varied from \(75~\mu\mathrm{m}\) to \(750~\mu\mathrm{m}\). This range of pinhole radii spans source spatial coherences from high to low (corresponding to small to large source radii), following the scaling used in the van Cittert-Zernike Schell formalism \cite{Montoya2025GVS}. Twenty-five distinct objects were simulated for each individual object class and source radius (1000 amplitude objects in total). Representative amplitude reconstructions are shown in Fig.~\ref{fig:representative_reconstructions}. Reconstructions of phase-objects, obtained using two detector planes, are further reported in the Supplementary Material. The coherence-aware model preserves the dominant object structure across the considered source radii. In contrast, reconstructions obtained under the fully coherent assumption become progressively less accurate as the source radius increases. The degradation is especially evident for dense objects, where the coherent model produces loss of fine structure, reduced feature contrast, and  reconstruction artefacts that are absent or substantially suppressed with the partially coherent  model. The consequences of coherence mismatch are thus seen to emerge directly in the recovered amplitude structure.

\begin{figure}[t]
	\centering
	\includegraphics[width=\linewidth]{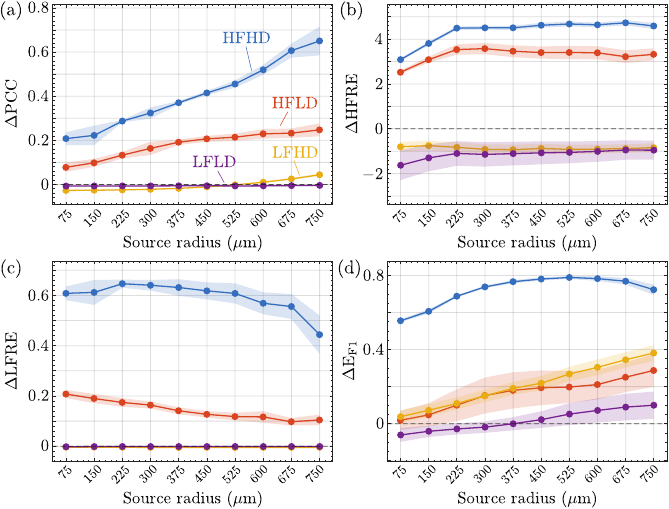}
	\caption{\textbf{Class-resolved reconstruction advantage under varying source coherence.}
		Differences $\Delta Q$ are plotted as a function of source radius for the four object classes. Panels show differences in (a) PCC, (b) HFRE, (c) LFRE, and (d) \(E_{F1}\). Solid markers denote the class-averaged value (across 25 distinct objects) at each source radius, and shaded bands denote standard deviation across objects within that class.}
	\label{fig:classwise_quantitative_trends}
\end{figure}

To quantitatively compare reconstruction quality we considered a number of distinct metrics \(Q_{\nu}\). First, the global reconstruction fidelity was quantified using the Pearson correlation coefficient (PCC) \cite{Pearson1895Correlation}. Next, to assess whether reconstruction preserved the spectral distribution of the reference object, the high-frequency retention error (HFRE) and low-frequency retention error (LFRE) were computed as the logarithmic mismatch between the reconstructed and reference spectral energy fractions above and below a fixed normalised radial-frequency cutoff of \(0.25\) cycles$/$pixel. Finally, boundary recovery was quantified using the edge \(F_1\) score, \(E_{F1}\), found from boundary precision and recall with a small spatial tolerance \cite{Canny1986EdgeDetection,Martin2004BoundaryDetection}. For each object class and source radius, reconstruction advantage was computed as the difference between quality metrics for the partially coherent and coherent forward models. Specifically, for metrics for which larger values indicate better reconstruction (\(E_{F1}\) and PCC), we set \(\Delta Q = Q_{\mathrm{pc}} - Q_{\mathrm{coh}}\), whereas for HFRE and LFRE, which are error measures, we set \(\Delta Q = Q_{\mathrm{coh}} - Q_{\mathrm{pc}}\). Positive values therefore always indicate that the partially coherent model provides improved image reconstruction. The resulting average reconstruction improvement scores (and corresponding standard deviation) for each object class and source radius are depicted in Fig.~\ref{fig:classwise_quantitative_trends}.

The results of Fig.~\ref{fig:classwise_quantitative_trends}  demonstrate a strong dependence on the spatial dimensionality and frequency content of the object class. HFHD objects exhibit the largest and most consistent positive improvement from use of $\mathcal{M}_{\text{pc}}$. Their dense, high-frequency structure generates many closely spaced boundaries and a broad spatial-frequency spectrum, making the recorded diffraction pattern highly sensitive to partial coherence. As the source radius increases, $\mathcal{M}_{\text{coh}}$ increasingly fails to represent this partial coherence-induced redistribution of spatial frequency content. The positive \(\Delta\mathrm{PCC}\) shows improved full-image reconstruction performance, while the positive \(\Delta\mathrm{HFRE}\) and \(\Delta\mathrm{LFRE}\) show that the partially coherent model better preserves the spectral distribution of the reference object. Positive \(\Delta E_{F1}\) further indicates more accurate boundary localisation. HFLD objects show a weaker, but still clear, improvement when using $\mathcal{M}_{\text{pc}}$. These objects contain fine features, but their lower spatial density reduces the number of independent high-frequency structures in the field. Consequently, \(\Delta\mathrm{PCC}\), \(\Delta\mathrm{HFRE}\), and \(\Delta E_{F1}\) remain positive, demonstrating improved reconstruction. 

LFHD amplitude objects show limited $\Delta$PCC, because their dominant structure lies at low spatial frequency and is therefore less affected by partial coherence. $\mathcal{M}_{\text{coh}}$ can capture much of the broad object envelope, so the difference between the partially coherent and coherent reconstructions remains small in global intensity and spectrum quality. Notably, the positive \(\Delta E_{F1}\) however shows that the $\mathcal{M}_{\text{pc}}$ still improves boundary localisation. The negative \(\Delta\mathrm{HFRE}\) arises since LFHD objects contain relatively little energy above the radial-frequency cutoff. In such a regime, small amounts of additional reconstructed edge sharpness or residual texture can increase the high-frequency spectral mismatch, even when $\mathcal{M}_{\text{pc}}$ improves boundary localisation. LFLD amplitude objects show the weakest improvement among all objects, as they and their corresponding reconstructions are dominated by broad slowly varying features. Full-image metrics are thus less sensitive to local boundary differences. The small or negative spatial and spectral differences, together with the weaker \(E_{F1}\) improvement, indicate that partial-coherence modelling produces only limited boundary-level gains for this class. 

Phase-object reconstructions show a related but more nuanced dependence on spatial coherence, as detailed in the Supplementary Material. Coherence-aware inversion improves boundary localisation across all phase-object classes, as indicated by consistently positive \(\Delta E_{F1}\). Global and spectral gains are however more strongly class dependent. The largest spectral improvements occur for high-frequency phase objects, whereas low-frequency phase objects can show weak or negative HFRE and LFRE changes despite improved boundary localisation.

\begin{figure}[t]
	\centering
	\includegraphics[width=\linewidth]{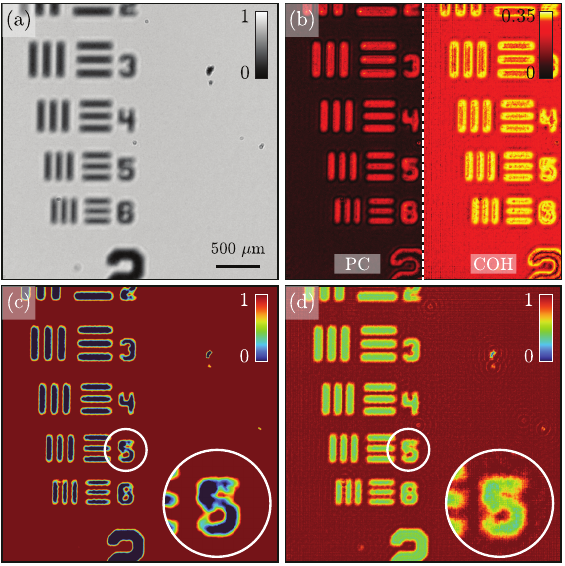}
	\caption{\textbf{Experimental verification using a USAF resolution target.}
		(a) Measured lensless detector intensity.
		(b) Absolute error map between the measured and predicted detector intensities, for partially (left) and fully (right) coherent models, respectively. Reconstructed target obtained using a (c) partially coherent and (d) fully coherent forward model. Insets showing magnified views of the indicated feature.
	}
	\label{fig:experimental_validation}
\end{figure}

Experimental verification was performed using a lensless measurement of a USAF resolution target under partially coherent illumination (Fig.~\ref{fig:experimental_validation}(a)). The target was illuminated by a commercial green light-emitting diode coupled through a \(500~\mu\mathrm{m}\) pinhole (Thorlabs P500HMB). The source-to-object and object-to-detector separations were \(z_{so}=110~\mathrm{mm}\) and \(z_{od}=2~\mathrm{mm}\), respectively. The diffraction pattern was recorded using a Raspberry Pi Camera Module 3 equipped with a Sony-megapixel array with a \(7.4~\mathrm{mm}\) sensor diagonal and a nominal pixel pitch of \(1.55~\mu\mathrm{m}\times1.55~\mu\mathrm{m}\). A central \(2000\times2000\)-pixel region of the recorded measurement was used for computational reconstruction. Figure~\ref{fig:experimental_validation} shows the reconstruction results obtained using both $\mathcal{M}_{\text{pc}}$ and $\mathcal{M}_{\text{coh}}$ operators. In the absence of a ground-truth object transmittance, the reconstruction quality was verified using the agreement between the measured and predicted detector intensities (Fig.~\ref{fig:experimental_validation}(b)), which was quantified using PCC and Fourier correlation (FC). The partially coherent model yielded \(\mathrm{PCC}=\mathrm{0.9871}\) and \(\mathrm{FC}=\mathrm{0.9906}\), compared with \(\mathrm{PCC}=\mathrm{0.9212}\) and \(\mathrm{FC}=\mathrm{0.9396}\) for the coherent model. The experimental results provide validation of the forward-model analysis and show that incorporating the source coherence improves agreement with the observed lensless intensity pattern.

In conclusion, our results establish spatial coherence as a consequential component of the lensless imaging inverse problem. Incorporating partial spatial coherence into the forward model improves reconstruction quality and yields closer agreement with measured detector intensities. The improvement is greatest when the target contains fine, densely distributed features, because such objects place a greater demand on accurate spatial-frequency transfer and boundary localisation. Our findings indicate that coherence should be treated as an explicit design and modelling parameter in lensless imaging, rather than as an incidental property of the illumination. Recent information-theoretic approaches to object-dependent lensless system design provide a natural framework for extending this analysis \cite{Kabuli2026InfoLensless}. Future work could use such criteria to optimise the source coherence for specified objects and imaging tasks, enabling object-adaptive lensless system design.

\begin{acknowledgments}
Ministry of Education, Singapore (RG137/24, EDUN C-33-18-279-V12); Agency for Science, Technology and Research, Singapore (H25-MRG3483).
\end{acknowledgments}

\end{document}

\title{Role of Spatial Coherence in Single-Shot Lensless Image Reconstruction: Supplemental Material}

\author{Ganesh M. Balasubramaniam$^1$\orcidlink{0000-0001-7824-5831}, Xiao-Liu Chu$^{2}$\orcidlink{0000-0002-7476-2097} and Matthew R. Foreman$^{1,2}$\orcidlink{0000-0001-5864-9636}}
\email[]{matthew.foreman@ntu.edu.sg}
\affiliation{
$^1$School of Electrical and Electronic Engineering, Nanyang Technological University, 50 Nanyang Avenue, Singapore 639798 \\
$^2$Institute for Digital Molecular Analytics and Science, 59 Nanyang Drive, Singapore 636921
}

\date{\today}

\maketitle
This supplemental document describes our object generation process and  extends the analysis of spatial coherence in lensless image reconstruction to phase objects. The analysis, therefore, considers unit-modulus phase transmittances with phase values ranging from \(0\) to \(2\pi\) and evaluates reconstructions from two detector-plane measurements, using the same partially coherent and coherent forward models as in the main article. Representative coherence-dependent measurements, phase reconstructions, and class-resolved quantitative results are presented for high-frequency high-density (HFHD), high-frequency low-density (HFLD), low-frequency high-density (LFHD), and low-frequency low-density (LFLD) objects. Reconstruction performance is assessed using the same metrics as in the main article, namely the Pearson correlation coefficient (PCC), high-frequency retention error (HFRE), low-frequency retention error (LFRE), and edge \(F_1\) score (\(E_{F1}\)). The results show that coherence-aware inversion improves phase boundary localisation most consistently, while global and spectral gains depend strongly on object structure and spatial frequency content. These findings provide supporting evidence that spatial coherence is a consequential component of the lensless inverse model for both amplitude and phase objects.

\section*{S1. Synthetic object image generation}

The synthetic object library used in this work was generated from glyph-based binary patterns using MATLAB \cite{lecun2010mnist}. Each object was created on a \(2000\times2000\) pixel grid and assigned to one of four structural classes, namely, high-frequency high-density (HFHD), high-frequency low-density (HFLD), low-frequency high-density (LFHD), and low-frequency low-density (LFLD). These classes were designed to independently vary the characteristic spatial scale of the object features and their areal density. High-frequency objects were generated using smaller, sharper glyphs, whereas low-frequency objects were generated using larger glyphs with Gaussian smoothing. High-density objects contained many glyphs distributed across the field of view, whereas low-density objects contained fewer glyphs.

For each class, glyphs were randomly selected from a digit or letter library, resized, rotated within a prescribed angular range, and placed onto the image canvas. An occupancy mask was used during placement to minimise glyph overlap and to enforce class-dependent spacing between neighbouring features. Grid-based placement was used for the high-density classes to distribute features broadly across the field, while random placement was used for the low-density classes. After placement, class-dependent Gaussian smoothing was applied to control the effective spatial-frequency content. The final amplitude objects were inverted to produce dark features on a bright background and normalised to the range \([0,1]\). A total of 25 objects were generated for each class.

Object-class separation was verified using a spatial-frequency score and a feature-density score. For each generated object \(I(\mathbf{r})\), the image was first normalised, mean-subtracted, and multiplied by a two-dimensional Hann window \(W(\mathbf{r})\). The centred Fourier power spectrum \(P(\mathbf{f})\) was then computed as
\begin{equation}
	P(\mathbf{f})
	=
	\left|
	\mathcal{F}
	\left\{
	\left[
	I(\mathbf{r})-\langle I(\mathbf{r})\rangle
	\right]
	W(\mathbf{r})
	\right\}
	\right|^{2}.
\end{equation}
A normalised radial spatial-frequency coordinate was defined as
\begin{equation}
	\rho(\mathbf{f})
	=
	\frac{
		\sqrt{(f_x-f_{x,0})^2+(f_y-f_{y,0})^2}
	}{
		\rho_{\max}
	},
	\qquad
	0 \leq \rho \leq 1 .
\end{equation}
The spatial-frequency score was computed as the fraction of spectral power above the radial-frequency cutoff \(\rho_c=0.18\),
\begin{equation}
	S_f
	=
	\frac{
		\sum_{\rho(\mathbf{f})\geq 0.18} P(\mathbf{f})
	}{
		\sum_{\rho(\mathbf{f})>0.015} P(\mathbf{f})
	} .
\end{equation}
The lower cutoff \(\rho>0.015\) removes the near-DC component from the normalisation, so that \(S_f\) reflects object structure rather than the average background level.

The feature-density score was computed from the binary object mask \(M(\mathbf{r})\) as
\begin{equation}
	S_d
	=
	\frac{1}{N^2}
	\sum_{\mathbf{r}} M(\mathbf{r}),
\end{equation}
where \(N=2000\) is the image size. The raw scores were robustly normalised across the generated library,
\begin{equation}
	\widetilde{S}
	=
	\mathrm{clip}
	\left[
	\frac{S-P_{2}(S)}
	{P_{98}(S)-P_{2}(S)},
	0,1
	\right],
\end{equation}
where \(P_2\) and \(P_{98}\) denote the 2nd and 98th percentiles, respectively. The four object classes were then verified in the \((\widetilde{S}_f,\widetilde{S}_d)\) plane using a cutoff of \(0.5\) along each axis. Thus, HFHD objects occupy the high-\(\widetilde{S}_f\), high-\(\widetilde{S}_d\) region, HFLD objects occupy the high-\(\widetilde{S}_f\), low-\(\widetilde{S}_d\) region, LFHD objects occupy the low-\(\widetilde{S}_f\), high-\(\widetilde{S}_d\) region, and LFLD objects occupy the low-\(\widetilde{S}_f\), low-\(\widetilde{S}_d\) region.

\section*{S2. Phase object reconstruction under partial spatial coherence}
\begin{figure}[!b]
	\centering
	\includegraphics[width=0.98\linewidth]{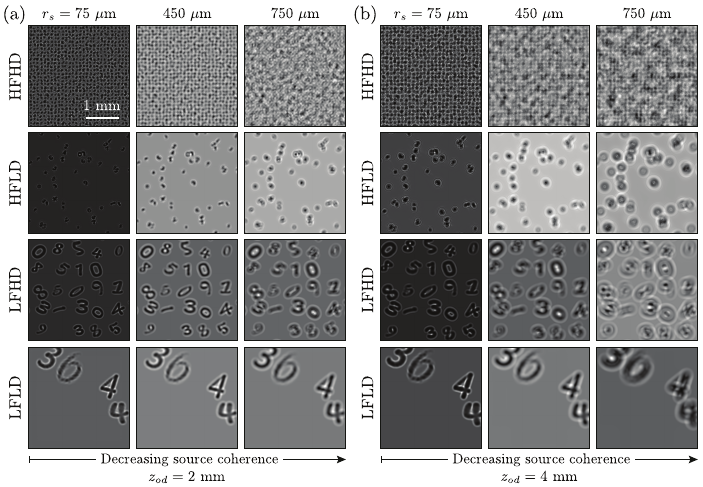}
	\caption{\textbf{Coherence-dependent lensless measurements for phase objects derived from the object classes in Fig.~1 of the main manuscript.}
		The same object classes used in Fig.~1 of the main manuscript are considered here as phase objects rather than amplitude objects. Rows correspond to high-frequency, high-density (HFHD), high-frequency, low-density (HFLD), low-frequency, high-density (LFHD), and low-frequency, low-density (LFLD) phase objects. The detector intensities correspond to lensless measurements under different source-coherence conditions. (a) and (b) show detector measurements at 2 mm and 4 mm, respectively.}
	\label{fig:supp_phase_image_formation}
\end{figure}
The phase object is modelled as a unit-modulus transmittance,
\begin{equation}
	t(\mathbf{r}_{o})
	=
	\exp\!\left[
	i\phi(\mathbf{r}_{o})
	\right],
	\qquad
	0 \leq \phi(\mathbf{r}_{o}) \leq 2\pi ,
	\label{eq:supp_phase_object}
\end{equation}
where \(\mathbf{r}_{o}=(x_o,y_o)\) denotes the transverse object-plane coordinate and \(\phi(\mathbf{r}_{o})\) is the spatially varying phase delay. The same partially coherent and coherent forward operators used in the main text were applied.

Phase recovery was performed using two detector-plane intensities, generated at \(z_{od}=2~\mathrm{mm}\) and \(z_{od}=4~\mathrm{mm}\). A single-phase estimate was propagated to both detector planes (see Fig.~\ref{fig:supp_phase_image_formation}), and the plane-resolved loss terms were combined into a common optimisation objective. This two-plane configuration provides additional propagation diversity, reducing ambiguities that are more pronounced with phase-only inversion. All other simulation parameters, including wavelength, source-to-object distance, source radii, and the comparison between partially coherent and coherent forward models, were kept identical to those used in the amplitude analysis in the main text.

\section*{S3. Representative phase reconstructions}

\begin{figure}[!htbp]
	\centering
	\includegraphics[width=\linewidth]{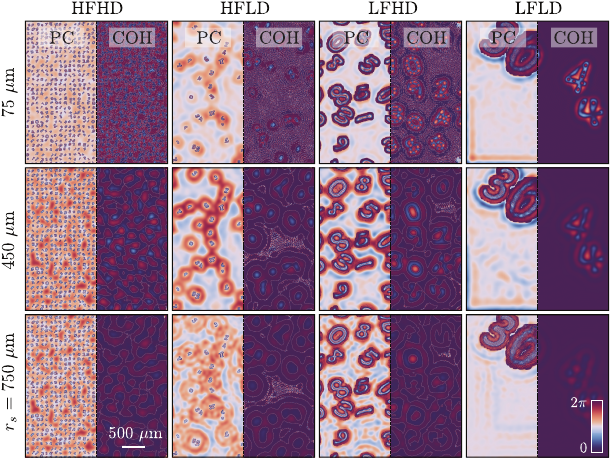}
	\caption{\textbf{Representative phase reconstruction comparison across object classes and source coherence.}
		Phase-object reconstructions for HFHD, HFLD, LFHD, and LFLD object classes. Rows correspond to source radii \(r_s=75~\mu\mathrm{m}\), \(450~\mu\mathrm{m}\), and \(750~\mu\mathrm{m}\). In each image, the left half shows the reconstruction obtained using the partially coherent (PC) forward model, while the right half shows the reconstruction obtained using the coherent (COH) forward model. The dashed line marks the boundary between the two reconstructions.}
	\label{fig:supp_phase_reconstructions}
\end{figure}

Representative phase reconstructions are shown in Fig.~\ref{fig:supp_phase_reconstructions}. In each panel, the left half corresponds to the reconstruction obtained using the partially coherent forward model, while the right half corresponds to the reconstruction obtained using the coherent forward model. The comparison shows that coherence mismatch also affects phase recovery, although the observed improvement differs from that in the amplitude case because phase reconstruction uses two detector planes.

For high-frequency phase objects, the partially coherent model more consistently preserves phase boundaries and local phase structure. In contrast, the coherent model can introduce background texture and oscillatory artefacts, particularly at larger source radii, because it attempts to explain a partially coherent measurement using a fully coherent propagation operator. For low-frequency phase objects, the difference between the two models is less uniform, because the dominant information is carried by broader spatial features and the two-plane configuration partially compensates for forward-model mismatch.

\section*{S4. Quantitative phase reconstruction analysis}

The quantitative phase analysis used the same four metrics as the main article. Global reconstruction fidelity was quantified using the Pearson correlation coefficient (PCC). Spectral preservation was quantified using high-frequency retention error (HFRE) and low-frequency retention error (LFRE), computed as the logarithmic mismatch between the reconstructed and reference spectral energy fractions above and below a fixed radial-frequency cutoff. Boundary recovery was quantified using the edge \(F_1\) score, \(E_{F1}\), computed from boundary precision and recall with a small spatial tolerance.

For metrics where larger values indicate better reconstruction, namely PCC and \(E_{F1}\), the reconstruction advantage was defined as
\begin{equation}
	\Delta Q
	=
	Q_{\mathrm{pc}}
	-
	Q_{\mathrm{coh}} ,
	\label{eq:supp_positive_delta}
\end{equation}
where \(Q_{\mathrm{pc}}\) and \(Q_{\mathrm{coh}}\) are the metric values obtained from the partially coherent and coherent forward models, respectively. For the error metrics HFRE and LFRE, the sign was reversed,
\begin{equation}
	\Delta Q
	=
	Q_{\mathrm{coh}}
	-
	Q_{\mathrm{pc}} .
	\label{eq:supp_error_delta}
\end{equation}
Thus, positive values always indicate an advantage for the partially coherent forward model.

\begin{figure}[!t]
	\centering
	\includegraphics[width=\linewidth]{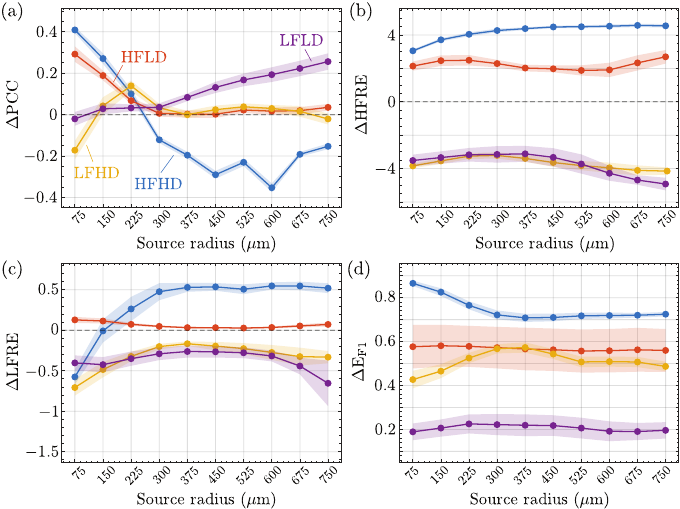}
	\caption{\textbf{Class-resolved reconstruction advantage under varying source coherence.}
		Paired reconstruction differences are plotted as a function of source radius for the four object classes. Panels show differences in (a) PCC, (b) HFRE, (c) LFRE, and (d) \(E_{F1}\). Solid markers denote the class-averaged value at each source radius, and shaded bands denote standard deviation across objects within that class.}
	\label{fig:supp_phase_metric_deltas}
\end{figure}

The phase reconstruction results in Fig.~\ref{fig:supp_phase_metric_deltas} show that coherence-aware inversion produces the most consistent improvement in boundary recovery. All four object classes show positive \(\Delta E_{F1}\) across the full range of source radii, indicating that the partially coherent forward model localises phase boundaries more accurately than the coherent forward model. The magnitude of this improvement depends strongly on object class. HFHD phase objects show the largest boundary advantage, with \(\Delta E_{F1}\) remaining high across all source radii. HFLD and LFHD objects show intermediate positive boundary gains, whereas LFLD objects show the weakest but still positive \(E_{F1}\) imrovement.

The PCC is more class-dependent. HFHD objects show a positive \(\Delta\mathrm{PCC}\) at small source radii, but this advantage decreases with increasing source radius and becomes negative at larger source radii. This indicates that the partially coherent model improves boundary localisation for dense, high-frequency phase objects, even when full-image correlation is no longer favourable. HFLD objects show a positive \(\Delta\mathrm{PCC}\) at small source radii and a near-zero improvement at larger source radii. LFHD objects show only a weak PCC advantage over most of the source-radius range. In contrast, LFLD objects show an increasing positive \(\Delta\mathrm{PCC}\) with source radius, indicating that the partially coherent model most clearly improves global phase recovery for sparse low-frequency phase objects under lower spatial coherence.

The spectral error metrics distinguish between high- and low-frequency phase classes. HFHD and HFLD objects show positive \(\Delta\mathrm{HFRE}\) across all source radii, demonstrating that the partially coherent model substantially reduces high-frequency spectral error for phase objects containing fine spatial structure. This effect is strongest for HFHD objects and remains large as the source radius increases. Conversely, LFHD and LFLD objects show negative \(\Delta\mathrm{HFRE}\), indicating that the coherent model gives lower high-frequency retention error for these low-frequency phase classes under this metric. This behaviour is expected because low-frequency objects contain limited reference high-frequency energy, so small differences in reconstructed edge sharpness or residual texture can dominate the high-frequency error measure.

The LFRE improvement shows a different pattern. HFHD objects show negative or near-zero \(\Delta\mathrm{LFRE}\) at small source radii, followed by positive values at larger source radii, indicating improved low-frequency spectral preservation by the partially coherent model once coherence mismatch becomes stronger. HFLD objects show a small but consistently positive \(\Delta\mathrm{LFRE}\). In contrast, LFHD and LFLD objects show negative \(\Delta\mathrm{LFRE}\) across the source-radius range, indicating that the coherent model gives lower low-frequency spectral error for these broader phase structures. The negative LFRE improvement is especially pronounced for LFLD objects at large source radii.

In summary, the phase analysis shows that partial-coherence modelling improves phase reconstruction in an object-dependent manner. The most consistent benefit is observed in boundary localisation, with positive \(\Delta E_{F1}\) across all phase-object classes and source radii. Spectral advantages are strongest for high-frequency phase objects, whereas low-frequency phase objects show weaker or negative spectral-error enhancement despite improvements in boundary recovery. These results support the conclusion of the main article that spatial coherence must be treated as an explicit component of the lensless inverse model, while showing that its effect on phase reconstruction depends on object structure and on the reconstruction property being evaluated.